\documentclass[manuscript]{emulateapj}

\def\gsim{\mathrel{\hbox{\rlap{\lower.55ex \hbox {$\sim$}}
                   \kern-.3em \raise.4ex \hbox{$>$}}}}
\def\lsim{\mathrel{\hbox{\rlap{\lower.55ex \hbox {$\sim$}}
                   \kern-.3em \raise.4ex \hbox{$<$}}}}

\def\he2{\hbox{He\,{\sc ii} $\lambda$4686}}

\def\RL1{\hbox{{$R_{L_{1}}$}}}

\shorttitle{Rotational Doppler beaming in eclipsing binaries}
\shortauthors{Paul J. Groot}

\begin{document}


\title{Rotational Doppler beaming in eclipsing binaries}


\author{Paul J. Groot\altaffilmark{1,2}}
\affil{Cahill Center for Astronomy and Astrophysics, California
  Institute of Technology, 1200 East California Boulevard, Pasadena, CA 91125}
\affil{Department of Astrophysics/IMAPP, Radboud University Nijmegen,
  P.O.Box 9010, 6500 GL, Nijmegen, The Netherlands} 
\email{pgroot@astro.ru.nl}


\begin{abstract}
In eclipsing binaries the stellar rotation of the two components will
cause a rotational Doppler beaming during eclipse ingress and egress
when only part of the eclipsed component is covered. For eclipsing
binaries with fast spinning components this photometric analogue of
the well-known spectroscopic Rossiter-McLaughlin effect can exceed the
strength of the orbital effect. Example light curves are shown for a
detached double white dwarf binary, a massive O-star binary and a
transiting exoplanet case, similar to WASP-33b. Inclusion of the
rotational Doppler beaming in eclipsing systems is a prerequisite for
deriving the correct stellar parameters from fitting high quality
photometric light curves and can be used to determine stellar
obliquities as well as e.g. an independent measure of the rotational
velocity in those systems that may be expected to be fully
synchronized. 
\end{abstract}





\section{Introduction}

Doppler beaming in stellar binaries was first discussed, in the
context of stellar rotation, by Hills \& Dale (1974), after which it
led a dormant life until it was rediscussed in the context of orbital
motion by Maxted, Marsh \& North (2000). The general theory was
extensively discussed in Loeb \& Gaudi (2003) and Zucker, Mazeh \&
Alexander (2007) for stellar binaries and star-planet cases in the
context of to the orbital motion of the components in a
binary. Because the flux density, F$_{\nu,0}$ at a frequency $\nu$ is
not a Lorentz-invariant quantity by itself, a brightening or dimming
of the observed flux occurs depending on the radial velocity of the
object.

In short, for non-relativistic velocities the
beamed flux, $F_\nu$ at frequency $\nu$ depends on the radial velocity
($v_r$) and spectral slope ($\alpha$) as (Eq. 2 from Loeb \& Gaudi,
2003):
\begin{equation}
F_\nu = F_{\nu,0} [ 1+ (3-\alpha)\frac{v_r}{c}] \equiv F_{\nu,0} B_{\alpha,v_r}, \label{eq:beaming}
\end{equation}
where $F_{\nu,0}$ is the unbeamed signal and $\alpha$ depends on the
observing frequency and spectral slope of the object. Eq.\ \ref{eq:beaming} defines the beaming factor $B_{\alpha,v_r}$. 
For blackbodies
at a temperature $T_{\rm eff}$ one can approximate $\alpha$ with Eq. 3
from Loeb \& Gaudi (2003) or Zucker et al. (2007):
\begin{equation}
\alpha(\nu) = \frac{e^x(3-x)-3}{e^x-1} = 3 - \frac{x e^x}{e^x-1},  \label{eq:alpha}
\end{equation}
with $x \equiv h \nu/kT_{\rm eff}$, and $h$ and $k$ Planck's and
  Boltzmann's constants, respectively. 

Observationally, Doppler beaming has been detected in a small number
of binaries so far. Van Kerkwijk et al. (2010) used the excellent
photometric quality of the {\sl Kepler} satellite to demonstrate the
existence of Doppler beaming in two binary systems consisting of an
A-type star and a white dwarf orbiting each other in 5.2 days and 23.9
days, respectively. Subsequently the same effect has been found in a
subdwarf B - white dwarf binary with a 9.6 hour orbital period by
Bloemen et al. (2010), also using {\sl Kepler} data, in the transiting
massive Jupiter/low-mass brown dwarf plus F3-type main sequence star
system CoroT-3b, with a 4.3 day orbital period by Mazeh \& Faigler (2010),
and in the eclipsing, detached double white dwarf systems NLTT\,11748
and SDSS\,J0651+38 in ground-based photometric data by Shporer et
al. (2010) and Brown et al. (2011). 

Apart from the Doppler beaming due to the orbital motion of the two
stars in a binary, the same effect will also occur due to the {\sl
  rotational} velocity of the stars. This was the original, but
incomplete, context of Hills \& Dale (1974) and it has also been
mentioned briefly by Van Kerkwijk et al. (2010).  As the star spins on
its rotational axis half of the star will be moving towards the
observer and half of it will be moving away from the observer. In
non-eclipse conditions and for axisymmetric stars the combined beaming
effect of these two halves exactly cancels. However, during eclipse
ingress and egress the partial coverage of the eclipsed component in
the binary will break the symmetry and a net beaming will be the
result. The amplitude of the beaming is set by the flux-weighted
effect of the non-eclipsed part of the star and will be simulated in
Section\ \ref{sec:simu} for three cases. A fourth case, of solar-type
binaries, has subsequently been discussed by Shporer et al. (2011). 

In radial velocity studies, the related shift in the observed radial
velocity of spectral lines during ingress and egress is well known as
the Rossiter-McLaughlin effect, as first shown by Rossiter (1924) and
McLaughlin (1924). The shape, amplitude and asymmetry of the
Rossiter-McLaughlin effect can be used to derive the projected
rotational velocity of the star, prograde or retrograde rotation and
the obliquity of the binary orbit with respect to the stellar
components. A detection of the rotational Doppler beaming (or
photometric Rossiter-McLaughlin effect) may open up a photometric way
of determining these same parameters in eclipsing binaries,
substituting high spectral resolution studies with high
signal-to-noise photometric time series. For very short period
binaries such as the recently reported 12 minute orbital period
detached white dwarf binary SDSS\,J0651+28 (Brown et al., 2011) the
effect can, in principle, also be used for an independent
determination of the spin-orbit coupling of the components to the
orbit.

\section{Rotational Doppler beaming}

For a spherical star with radius $R$ the radial component $(v_r)$ of
the rotational velocity $v_{\rm rot}$ simply scales as 
\begin{equation}
v_r = v_{\rm rot} (\frac{x}{R}) \sin i,
\end{equation}
with $x,y,z$ a right-handed coordinate system with $x$,$y$ defining
the orbital plane, and $z$ in the direction of orbital angular
momentum for an edge-on system. In our sign convention prograde
rotation of the stars with respect to the orbit results in positive
rotational velocities. The orbital inclination with respect to the
plane of the sky, $i$, is defined, as usual, to be 0\degr\ for a
face-on orbit and 90\degr\ for an edge-on orbit.

The maximum rotational velocity of a star is set by its break-up
velocity, $v_{\rm break} = \sqrt{\frac{MG}{R}}$, and its minimum
velocity is in principle unbounded, but, in short-period systems where
tidal forces are strong enough to synchronize the system, will be set
by the synchronisation velocity $v_{\rm sync}= \frac{2 \pi R}{P_{\rm
    orb}}$. In wide binaries where tidal synchronization is
ineffective, there is no hard lower limit to the rotational velocity
of each of the components.  In these equations $M$ is the mass of one
of the components of the binary, $R$ the radius of this component,
$P_{\rm orb}$ the binary orbital period and $G$ Newton's gravitational
constant. For main-sequence stars maximum rotational velocities are
reached for early-type stars and can exceed 300 km s$^{-1}$. After the
onset of the magnetic dynamo around spectral type F1V, when stars
develop a convective outer layer, the rotational velocity drops very
quickly due to magnetic braking (see e.g. Groot, Piters \& Van
Paradijs, 1996). In white dwarfs very high rotational velocities can
be reached. Using the Eggleton mass-radius relation for fully
degenerate white dwarfs as reported by Verbunt \& Rappaport (1988), a
1.2 M$_\odot$ white dwarf can have a break-up velocity of $>$6\,000 km
s$^{-1}$.

\begin{figure*}
\includegraphics[angle=-90,width=14cm]{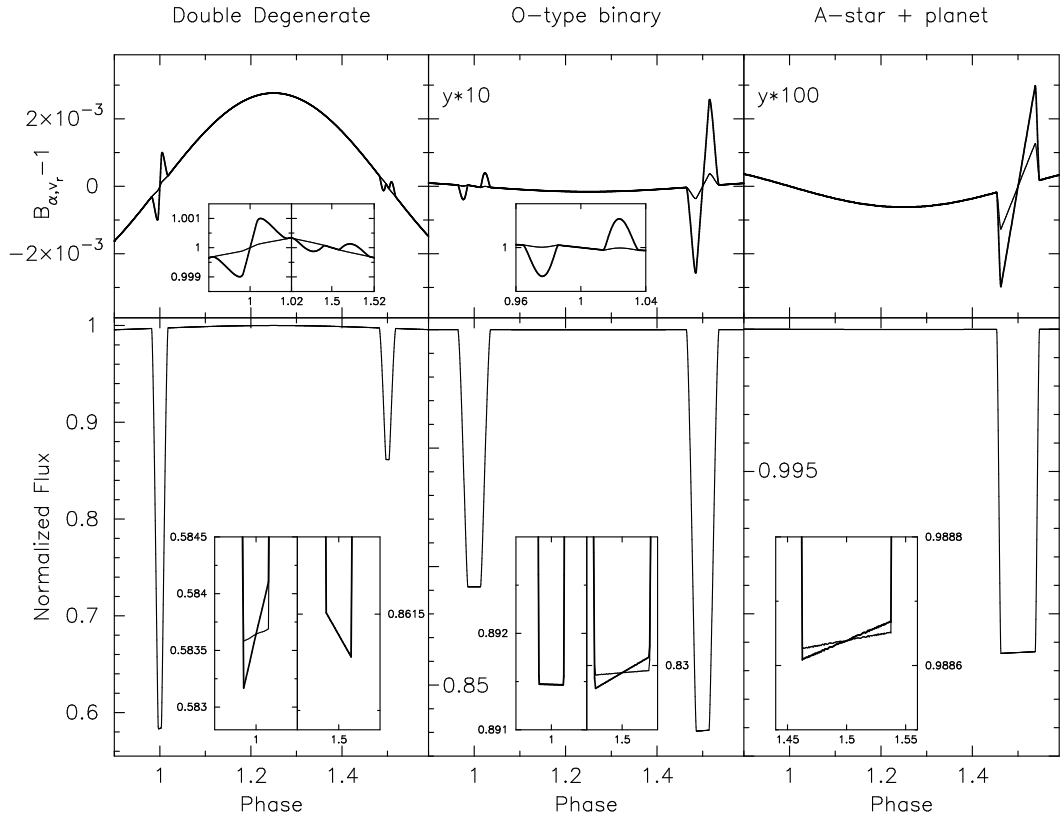}
\caption{Simulated eclipses ({\sl bottom}) and the photometric
  Doppler beaming factor ({\sl B$_{\rm \alpha, v_r}$, top}) for three eclipsing
  binary systems: a detached double degenerate system ({\sl
    left}), a massive O-type binary ({\sl middle}) and a planet
  around an A-star ({\sl right}). Inserts show close-ups of the light
  curve and beaming profile during the eclipses. Thick lines are for
  $v_{\rm rot} = 0.5v_{\rm break}$ and thin lines are for $v_{\rm
    rot} = v_{\rm sync}$, except for the star-planet case where the
  thin line is for v$_{\rm rot}$ = 86.5 km s$^{-1}$. \label{fig:simu}}
\end{figure*}

\begin{table}
\begin{center}
\caption{System parameters used in the simulations shown in
  Fig.\ \ref{fig:simu}}
\label{tab:system}
\begin{tabular}{llll}
\tableline\tableline
Parameter & double degenerate & O/B star & A-star+planet
\\
\tableline
M$_1$ ($M_\odot$)   &0.43  & 80  & 1.495 \\
M$_2$ ($M_\odot$)   &0.17  & 20  & 0.00392 \\
R$_1$ ($R_\odot$)   &0.0148  & 14 & 1.44 \\
R$_2$ ($R_\odot$)   &0.0214  & 6.1 &0.153  \\
T$_1$ ($K$)          &10\,000 &47\,800 &6\,440 \\
T$_2$ ($K$)          &16\,485 &34\,200 &1\,657 \\
P$_{\rm orb}$ ($d$)    &0.0027  &10.00  &1.22  \\
$i$ ($\degr$)& 90.0     & 90.0 & 90.0 \\
K$_1$ ($km\, s^{-1}$)  &218.8  &9.2  & 0.5 \\
K$_2$ ($km\, s^{-1}$)  &553.5  &36.7  &185.9  \\
v$_{\rm rot,1}$ ($km\, s^{-1}$) &1\,148.8 \& 27.7  & 510.0 \& 70.9  & 217.4 \& 86.4  \\
v$_{\rm rot,2}$ ($km\, s^{-1}$) & 601.3 \& 40.0  & 386.3 \& 30.9  &  34.1 \& 6.4 \\
$\alpha_1$ & 0.36 &1.72 &--0.81 \\
$\alpha_2$ & 1.10 &1.61 &--11.47 \\
\tableline
\end{tabular}
\end{center}
\end{table}

\section{Simulations} \label{sec:simu}
 To illustrate the rotational beaming effect we used Kepler's laws,
 the white dwarf mass-radius relation mentioned above, a massive star
 main-sequence mass-radius relation (at solar metallicity) from Pols
 et al.  (1998), stars with no limb darkening (but, see
 Sect.\ \ref{sec:limbdarkening}) and Eqs.\ \ref{eq:beaming} \&
 \ref{eq:alpha}. The two stars are denoted by $M_1$ and $M_2$ with
 $M_1$ the more massive object, and the mass ratio $q = M_2/M_1 <
 1$. Each star is given a blackbody spectrum characterized by the
 effective temperature $T_{\rm eff}$. The orbital Doppler beaming has
 been included in the simulations, but, for clarity and simplicity,
 additional effects such as ellipsoidal variations, the reflection
 effect and gravitational lensing have not been included. For
 illustration purposes all binaries are assumed to be seen exactly
 edge-on ($i$=90\degr).  Three types of binaries have been simulated
 and an overview of system parameters is given in
 Table\ \ref{tab:system}: a detached white dwarf - white dwarf binary
 similar to the one recently reported recently by Kilic et al. (2011)
 at an orbital period of 39 minutes; a massive O-star binary
 consisting of two main sequence stars of 20 M$_\odot$ and 80
 M$_\odot$ in a 10 day period orbit; and a transiting exoplanet around
 a fast rotating A-type star, similar to the WASP-33b system (Collier
 Cameron et al., 2010).  For each system the effect is shown for an
 effective wavelength of $\lambda$=6000 \AA, which, together with the
 effective temperature, determines the $\alpha$ factor in
 Eq.\ \ref{eq:alpha}. In all three systems the rotational velocity of
 both components is varied between break-up and synchronization. For
 the WASP-33b-like exoplanet case, the measured rotational velocity of
 the A-star ($v_{\rm rot}\sin i = 86.5$ km s$^{-1}$) was also
 simulated. The orbital phase $\phi$ is defined with respect to
 superior conjunction of the secondary. The projected surface area of
 the components has been divided in a 2\,000$\times$2\,000 grid with
 dimensions of $2R \times 2R$, where positive values for the flux and
 rotational velocity were assigned for all values within a radius
 equal to $R$, and all points outside $R$ were set to zero. Fluxes
 were assigned as relative to the other component in the binary, with
 the flux of the secondary component set to unity. The position of the
 two components along the orbit was calculated for 4\,000 phase
 bins. At each phase bin the projected areal overlap between the two
 components has been calculated and the flux from the eclipsed
 component adjusted accordingly. A flux-weighted mean rotational
 velocity of the star is calculated, and checked to be consistent with
 zero (within the numerical noise and finite grid width) for all
 phases outside of the primary or secondary eclipse.

\begin{figure}
\includegraphics[angle=-90,width=8.4cm]{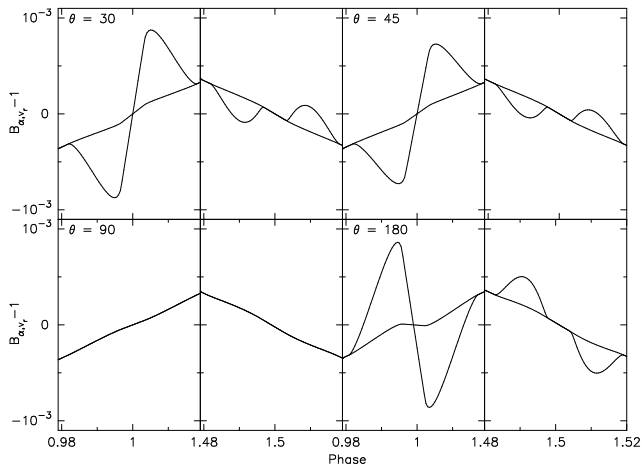}
\caption{Beaming factors for a detached double degenerate system as
  in Fig.\ \ref{fig:simu}, but now with both components obliquely
  rotating with respect to the orbital plane at an angle $\theta$, as
  indicated in the four sub panels. \label{fig:oblique}}
\end{figure}

Figure\ \ref{fig:simu} shows the results for the three binaries in
each of the three panels. The top panels shows the total Doppler
beaming factor $B_{\alpha, v_r}$ for both components in the
binary. For clarity both the orbital effect (the slow sinusoidal
variation with a period equal to the orbital period) as well as the
rotational effect (the excursions on top of the orbital variations
during eclipse ingress and egress) are shown. In each
figure the thick line indicates both components rotating at $v_{\rm
  rot} = 0.5 v_{\rm break}$ and the thin line has both
components rotating at $v_{\rm rot} = v_{\rm sync}$ with the exception
of the panel on the transiting planet, where the star has been set to
rotate at the measured rotational velocity of $v_{\rm rot} = 86.5$ km
s$^{-1}$. Fig.\ \ref{fig:simu} shows that for substantial rotational
velocity of the luminous component in the binary, the rotational
Doppler beaming effect can be as large or larger than the orbital
variation, similar to the spectroscopic Rossiter-McLaughlin effect
(Gaudi \& Winn, 2007). It is also clear from Fig.\ \ref{fig:simu} that a correct
inclusion of the rotational Doppler beaming is a prerequisite is to
derive the correct parameters of the components in the binary from
eclipse light curve fitting when very high quality data on a promising
system is available.

\begin{figure}
\includegraphics[angle=-90,width=8.4cm]{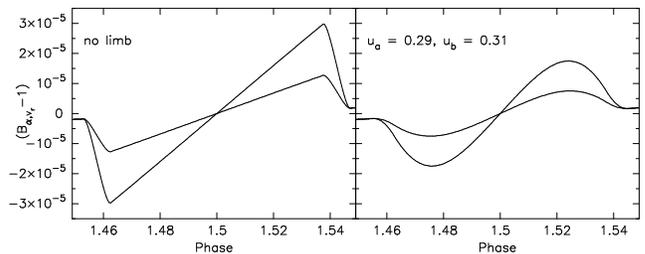}
\caption{Beaming factors for a transiting extrasolar planet as in
  Fig.\ \ref{fig:simu},
  not-including any limb-darkening {\sl (left)} and realistic limb
  darkening using a quadratic limb darkening law with coefficients
  $u_a$ and $u_b$ {\sl (right)}. Rotational velocities of the A-star
  primary are as given in Tab.\ \ref{tab:system}: 217.4 km s$^{-1}$
  (thick line) and 86.4 km s$^{-1}$ (thin line).  
\label{fig:limb}}
\end{figure}

\subsection{Obliquity and limb darkening} \label{sec:limbdarkening}

The shape and amplitude of the rotational Doppler beaming effect not
only depends on the rotational velocity. Various other effects will
play a role in determining the exact shape: the inclination of the
orbit and the obliquity of the rotational axis, oblateness and axial
symmetry of the stars, limb-darkening and/or limb-brightnening, and
differential rotation. In non-eclipsing, but highly distorted systems,
where, e.g. one of the components is (nearly) filling its Roche lobe
and is for instance irradiated asymmetrically by an accretion disk hot
spot, the asymmetric distribution of light on the distorted star can
even lead to ellipsoidal terms on the orbital signal. For the moment,
however, these highly distorted systems will not be considered and
only an illustration of the obliquity and limb darkening will be
given.

The  same double  degenerate  binary as  in  Sect.\ \ref{sec:simu}  is
taken, but  now the obliqueness,  $\theta$, of the rotational  axis of
the  two stars with  respect to  the orbital  plane is  varied between
$\theta$   =  30\degr,  45\degr,   90\degr\,  (oblique   rotator)  and
180\degr\,   (retrograde  rotator).   As  expected,   the   effect  of
obliqueness  of the  rotational  axis is  to  decrease and  eventually
reverse the  amplitude of the  beaming. If a  star is precessing  in a
binary  this  will  present  itself  as a  changing  obliqueness  with
time.  An  observational  detection  through  the  rotational  Doppler
beaming  would allow e.g.  a determination  of the  spin-orbit angular
momentum coupling in close binaries and/or hierarchical triples.

Limb darkening is a fundamental property of stellar atmospheres which
is reasonably well understood in main-sequence stars, but not so much
in white dwarf atmospheres. In helium dominated, high density white
dwarf atmospheres limb darkening could even be largely absent due to
internal refraction within the atmosphere (Kowalski \& Saumon, 2004).

Using a quadratic limb darkening law, 
\begin{equation}
F_\nu(\mu) = F_0 [1 - u_a(1-\mu) - u_b(1-\mu)^2], \label{eq:limb}
\end{equation} 
with $\mu = \cos\xi$ and $\xi$ the angle between the normal to the
surface and the line of sight of the observer to describe the limb
darkening, Fig.\ \ref{fig:limb} shows the effect for the transiting
planet case (Fig.\ \ref{fig:simu}), for limb darkening coefficients of
$u_a$=0.29 and $u_b$=0.31, appropriate for a $T_{\rm eff}$ = 6\,440 K
star in the {\sl Kepler} pass band (Sing 2010). The general effect of
limb darkening is to `soften' the edges of a star, from which it is
conceptually easy to deduce that it will impact the height of the
rotational Doppler beaming, in the same way that it affects the
Rossiter-McLaughlin effect. The contribution to the total light from
the most extremely beamed parts of the stars is lessened for a positive
limb darkening. Fig.\ \ref{fig:limb} shows that indeed not
only the amplitude of the Doppler beaming is supressed, but also the
skewness of the profile is significantly reduced. It is clear from
this figure that a realistic modeling of transiting planet light
curves where the Doppler beaming is included, needs to include a limb
darkening term as well. However, if indeed, in high density, helium
dominated white dwarf atmospheres the limb darkening coefficient is
close to zero, the resulting rotational Doppler beaming should hardly
be affected and may still show the sharp edges displayed in
Fig.\ \ref{fig:simu}.

\section{Detectability}

The ultimate goal of these investigations would of course be to first
detect and subsequently use the rotational Doppler beaming to
determine and constrain stellar and binary parameters. As can be seen
from Fig.\ \ref{fig:simu} even in the best of cases, such as the
double degenerates, this detection is going to be challenging. The
fractional amplitude of the rotational beaming varies between 10$^{-3}
\lsim B_{\alpha, v_r} \lsim 10^{-6}$. To firmly establish the shape
and amplitude the photometric accuracy needed would be a factor 10
smaller than this amplitude at least. The {\sl Kepler} light curves
show precisions in the 10$^{-4}$ - 10$^{-5}$ range, and it should
therefore be possible to detect the effect. In fact, a close look at
Fig.\, 3 in Bloemen et al. (2010) appears to show a residual effect in
the primary eclipse profile, by comparing their `R.+E.+L.+Beaming'
panel with the inset in the left panel of Fig.\ \ref{fig:simu} in
Sect.\ \ref{sec:simu}. However, this `by eye' residual, does not
clearly show up in the `Residuals' panel in Fig.\, 3 in Bloemen et
al. (2010).

In general the detectability is a combination of source flux, eclipse
duration, instrument stability and intrinsic amplitude. Source flux
can be optimized using a bigger telescope, eclipse duration is
unfortunately a given for an individual source, instrument stability
is in the design of the instrument and tends to be highest for space
missions, even though modern photometers on the ground now achieve
{\sl Kepler}-like precision (e.g. de Mooij et al., 2011). For a more
formal derivation of detectability I refer to Shporer et al. (2011),
but the recently detected, eclipsing 12 minute orbital period binary
SDSS\,J0651+28 (Brown et al., 2011) makes for an illustrative
example. Using the stellar parameters listed in Brown et al. (2011)
and assuming a fully edge-on, aligned and synchronized system (but see
Piro 2011), the fractional amplitude of the Doppler beaming would be
1.3$\times$10$^{-3}$. The best way to determine the rotational beaming
signal is to determine the light curve shape during primary
mid-eclipse (inset of Fig.\ \ref{fig:simu}), which for SDSS\,J0651+28
lasts for $\sim$0.05P$_{\rm orb}$. As a rule of thumb I will use a
minimum of 10 samples per eclipse and a photometric precision of a
tenth of the amplitude to be a minimum for a positive detection of the
effect, i.e. for SDSS\,J0651+28 a precision of 10$^{-4}$ in the
photometry needs to be obtained in 4 seconds on a $g=19.06$ magnitude
object. Assuming only Poisson statistics as a noise-contributor one
therefore needs to collect 10$^{8}$ photons per 4 seconds interval,
requiring the equivalent of a 180-meter telescope, even far out of
reach in the ELT era\footnote{Here I have used a photon flux of 10$^6$
  photons s$^{-1}$ cm$^{-2}$ for a $g$=0 object at the top of the
  Earth's atmosphere}. Of course, assuming perfect stability in the
source, one can stack an arbitrary number of orbits to achieve the
same precision. Even on a 10-meter class telescope this would require
the stacking of more than a thousand orbits, which is unfeasible in
the short run, unless we find a system that is 5-6 magnitudes brighter
in the sky than SDSS\,J0651+28.

\medskip
A much more promising case is presented by the O-type binaries, where
the orbital period is much longer (10 days in our example), and,
although mid-eclipse lasts for a shorter part of the orbital period,
roughly 0.025P$_{\rm orb}$, the `tenth of mid-eclipse' requirement
translates into 36 minutes during which to collect the signal. The
fractional amplitude is now only $10^{-4}$, requiring a precision at the
10$^{-5}$ level, but {\sl Kepler} has shown this to be
feasible. We now need to collect the photons at a rate of
4.5$\times$10$^{6}$ per second.  If our hypothetical binary were to
be $g$=14.0, this is achievable on a 4-m class telescope by stacking a
modest number of orbits (order 10), again assuming perfect source and
instrument stability. In practice of course, atmospheric and
electronic (detector) stability may very well be the limiting factor. 

\section{Discussion and conclusion}

For very high quality data on eclipsing systems inclusion of the
effect of rotational Doppler beaming in light curve fitting procedures
is necessary to derive the correct parameters of the binary. Since
Doppler beaming is a geometrical effect, not including it will lead
to a systematic error in the parameter derivation. The sign and
magnitude of the error depends very much on the system properties of
the binary. In prograde rotating binaries the extra `dimming' of the
binary at ingress and the extra `brightnening' during egress will
cause an asymmetric eclipse profile that will skew a timing solution
to a mid-eclipse phase that is too early. For retrograde rotation the
effect on the timing will be the opposite.

From the simulations shown here it is clear that the detection of
rotational Doppler beaming should be a mere matter of time and depends
on the availability of suitable systems more than on the accuracy of
current photometric systems. The most promising case would be a
non-synchronously rotating eclipsing double degenerate system, if it
is bright enough ($V<15$). Currently none of these are known.
Observationally there is a strong bias to find eclipsing double
degenerates in short-period systems where the ratio of the white dwarf
radius over the orbital separation is smallest and the system will be
observed to be eclipsing over a wider range of inclinations.
Unfortunately short-period double-detached binaries also have the
highest probability of (nearly) synchronous rotation since they are the product
of two episodes of common-envelope evolution (e.g. Nelemans et al.,
2001). This assumption does, however, depend strongly on the
rotational coupling of the core of a red giant to its outer envelope
(Sweigart \& Mengel, 1979), and on the time scale of the
common-envelope phase itself compared to the synchronisation time
scale (see e.g. Taam \& Sandquist, 2000).  Since the common-envelope
phase is expected to be relatively short, a decoupled core may remain
unsynchronized. Prime targets should therefore be young, hot, white
dwarfs, or even cores of planetary nuclei, in short period systems
with another white dwarf, i.e. just after the `birth' of the second
white dwarf in the system. It is unfortunate that the central binaries
in the nova - planetary nebula system V458 Vul (Wesson et al., 2009;
Rodr{\'\i}guez-Gil et al., 2010) and the planetary nebula system Hen
2--248 (Santander-Garc{\'\i}a et al., 2010) do not appear to be
eclipsing as these would have been almost ideal systems. The eclipsing
double degenerate system NLT\, 11448 (Steinfadt et al. 2010) is also
an obvious candidate for detection of rotational Doppler
beaming. Although promising at first we have shown that the faintness
adn short period of SDSS\,J0651+28 makes it a much harder target. 

Although the expected effect is a factor 10 smaller than in double
degenerate systems, eclipsing O/B stars are perhaps an even more
promising class of candidates. In general the orbital periods will be
much longer (days instead of hours), which causes the eclipse ingress
and egress to last much longer, and therefore also allows for the
accumulation of very high signal-to-noise ratio data. Since accurate
photometry requires the presence of many comparison stars to obtain
high precision differential light curves, slightly fainter, far-away
and/or reddened O/B binaries in the magnitude range 10$<V<$15 may be
more attractive candidates than very bright (V$<$10) nearby systems.

\end{document}